Research Paper

# Top research performance in Poland over three decades: A multidimensional micro-data approach

Marek Kwiek[a,b,*], Wojciech Roszka[a,c]

[a] Center for Public Policy Studies, Adam Mickiewicz University, Poznan, Poland
[b] German Center for Higher Education Research and Science Studies (DZHW), Berlin, Germany
[c] Poznan University of Economics and Business, Poznan, Poland

ABSTRACT

In this research, the contributions of a highly productive minority of scientists to the national Polish research output over the past three decades (1992–2021) is explored. A large population of all internationally visible Polish scientists ($N$ = 152,043) with their 587,558 articles is studied. In almost all previous research, the approaches to high research productivity are missing the time component. Cross-sectional studies were not complemented by longitudinal studies: Scientists comprising the classes of top performers have not been tracked over time. Three classes of top performers (the upper 1 %, 5 %, and 10 %) are examined, and a surprising temporal stability of productivity patterns is found. The 1/10 and 10/50 rules consistently apply across the three decades: The upper 1 % of scientists, on average, account for 10 % of the national output, and the upper 10 % account for almost 50 % of total output, with significant disciplinary variations. The Relative Presence Index (RPI) we constructed shows that men are overrepresented and women underrepresented in all top performers classes. Top performers are studied longitudinally through their detailed publishing histories, with micro-data coming from the raw Scopus dataset. Econometric models identify the three most important predictors that change the odds ratio estimates of membership in the top performance classes: gender, academic age, and research collaboration. The downward trend in fixed effects over successive six-year periods indicates increasing competition in Polish academia.

## 1. Introduction

The total national research output is produced by scientists from different disciplines and institutions, by men and women of different academic seniorities, biological ages, levels of research experience, and collaboration and publishing patterns. However, most importantly, the national research output is produced by scientists with highly skewed individual publishing productivity (David, 1994): Over the course of their careers, a small minority of scientists author large numbers of publications, and most scientists publish little (Crane, 1965; Allison, 1980; Xie, 2014; Ruiz-Castillo & Costas, 2014), with individual performance following a Paretian (power law) rather than Gaussian (normal) distribution (O'Boyle and Aguinis, 2012). The science profession is vertically stratified, and the skewness of science has been widely discussed in the past half century, mostly in bibliometric and economics of science literature (e.g., Cole & Cole, 1973; Seglen, 1992; Stephan, 2012).

* Corresponding author.
*E-mail addresses:* kwiekm@amu.edu.pl (M. Kwiek), wojciech.roszka@ue.poznan.pl (W. Roszka).

In the present paper, we explore the contribution of the highly productive minority of scientists (termed the top performers class) to the national Polish research output over the past 30 years (1992–2021). The contribution of top performers will be studied longitudinally at the micro-level of individual scientists. Details of their individual publishing histories (from the Scopus database) are unpacked and subsequently aggregated to higher levels of academic disciplines, institutional types, academic experience groups, and six-year periods of time—at which top performers (the upper 1 %, 5 %, and 10 %) are contrasted with the rest of the Polish scientific workforce (the remaining 99 %, 95 % and 90 %). We conceptualize top performers based on their output relative to the output of others.

Top performers have always produced the bulk of publications in all national science systems (Rosen, 1981; Xie, 2014; Ioannidis et al., 2014). However, it is interesting to analyze their role in a single, hugely expanding science system (Poland) from both temporal and cross-disciplinary perspectives. In other words, our interest is whether, during the 30 years studied, the role of top performers in academic knowledge production has remained unchanged amid intensive socio-economic changes accompanied by radical transformations in higher education and science systems.

Our approach is quantitative, and we study 152,043 Polish scientists from 15 STEMM disciplines (science, technology, engineering, mathematics, and medicine) and their 587,558 research articles published in five 6-year periods (1992–2021) in globally indexed Scopus journals. Our focus is on scientists and their properties rather than publications and their attributes. Our classes of top performers are defined using publication and citation metadata so that individual publishing productivity is normalized to the six-year period, discipline, and journal prestige.

### 1.1. *The focus on top research performance*

Top performance—from software engineers to entertainers, musicians, comedians, and professional athletes—attracts scholarly attention (O'Boyle & Aguinis, 2012; Rosen, 1981). The economics of superstars (Rosen, 1981) explains the skewness toward the most talented people in these activities. Top research performance (both in terms of publications and impact) is not an exception, attracting the attention of academic and science policy communities at all levels, from departments to institutions to national systems (Aguinis & O'Boyle, 2014).

Exceptionally high producers of scholarly publications, or "the prolific," are "strategic" in studying academic science and examining academic careers, being a window into analyzing the social stratification of science (Fox & Nikivincze, 2021). The entire traditional architecture of the academic enterprise, as presented in the traditional sociology of science and sociology of academic careers (Hermanowicz, 2012), is based on publications, their impact, and individual publishing productivity (or publications within a unit of time). Both recognition in science and academic progression, promotion, tenure, research funding, and access to collaboration networks are publication and impact driven (Stephan, 2012). There are two main classes of drivers behind publishing productivity: individual and environmental drivers (which include both departmental and institutional "work climates" and national science systems with different recognition, incentive, and promotion arrangements) (see Fox & Mohapatra, 2007; Leisyte & Dee, 2012; Tang & Horta, 2023).

As visible minorities, highly productive scientists have been examined through both qualitative and quantitative approaches. The qualitative approach included interviewing scientists listed as top performers to analyze why they are so productive (e.g., Kiewra & Creswell, 2000; Mayrath, 2008), and the quantitative approach included survey-based and bibliometric-based studies. Both have their own limitations, which are widely studied in the literature; however, they are complementary in analyzing the phenomenon of high productivity in science.

The top performers have been discussed over the past few decades under different labels, including stars, superstars, and star performers (as in Serenko et al., 2011; Aguinis & O'Boyle, 2014; Agrawal et al., 2017); the prolific and prolific professors (as in Fox & Nikivincze, 2021; Piro et al., 2016); top producing authors (Halevi et al., 2015); hyperprolific authors (Ioannidis et al. 2018); as well as top researchers (Cortés et al., 2016) and scientific elites (Larivière et al., 2010; Parker et al., 2010; Kwiek, 2016). The classes of scientists examined in the high-productivity literature have had a single attribute in common: They were all at the right end of the publishing productivity and impact distributions within universities, disciplines, countries, and globally.

Some analyses of top performance have focused on publications in refereed journals, without direct links to citations (to publication- or journal-level impact metrics)—but with rich survey data about both the social characteristics of scientists and features of their departments (see work habits and lifestyle choices in Parker et al. (2010), departmental climates in Fox & Nikivincze (2021), and academic beliefs and behaviors in Kwiek (2018a)); others focused on publications normalized to their location in national science systems as expressed in "publication points" (e.g., Piro et al. (2016) who examined the impact of the most productive Norwegian professors on their departments' citation impact; and Kyvik (1989) who studied publishing inequality between the various fields of science in Norway).

Still others have used publications normalized to journals, including "top journals" identified using journal impact factors (O'Boyle & Aguinis, 2012) and classes of journals weighted with different "quality points" (Carrasco & Ruiz-Castillo, 2012). Publications in top journals in respective disciplines are believed to be the most important antecedents of meaningful outcomes for faculty, including salary and tenure status (O'Boyle & Aguinis, 2012). Gomez-Mejia and Balkin (1992)) developed the conception of "prestige maximization" by scientists for themselves and their institutions, in which publications in top journals bring much more prestige than in any other outlets. The intrinsic value of papers was empirically found (Larivière & Gingras, 2010) as not being the only reason why papers are cited or not: a specific Matthew effect attached to journals is reported, apart from the Matthew effect attached to individual scientists. Larivière and Gingras (2010) found 4532 pairs of identical papers (duplicates) published in two journals and demonstrated that the same papers published in high-impact factor journals obtained twice as many citations as their identical counterparts in lower-impact factor journals.

The identification of top performers in the literature includes various methods of computing productivity: journal article numbers (Fox & Nikinincze, 2021); journal article numbers normalized to journal-level prestige (Abramo et al., 2009; O'Boyle & Aguinis, 2012); different publication types recalculated as article equivalents (Kyvik, 1989; Kwiek, 2018a); citations normalized to disciplines (e.g., Parker et al., 2010); as well as the various combinations of publication and citation-based impact indices (Sidiropoulos et al., 2016). In their study of "the skyline operator," in which a long list of carefully clustered indicators of high performance was used, Sidiropoulos et al. (2016) analyzed high performance as a multidimensional construct: In this conception, a very tiny minority of scientists cannot be surpassed by others with respect to all attributes at the same time, winning in each and every high-performance index; the resulting set of distinguished individuals is called "the skyline set."

The steep productivity stratification of scientists has been examined in both academic career studies (mostly through the datasets of academic profession surveys) and bibliometrics (mostly through Web of Science or WoS, Google Scholar, Microsoft Academic Search or MAS, and, more recently, Scopus datasets). In his survey-based study of Norwegian scientists, Kyvik (1989) found out that, no matter how productivity is calculated, the most prolific 20 % of scientists across various fields of science produced about 50 % of the national Norwegian research output. Similarly, Abramo et al. (2009) showed that 12 % of Italian scientists accounted for 35 % of the national Italian research output, with substantial cross-disciplinary variations.

The subsamples of top performers used in major studies usually do not exceed 20 % of the sample being examined: top performers examined are the upper 20 % of Norwegian scientists based on publication points (and the number of WoS publications) in Piro et al. (2016); top performers are defined as the upper 20 % of economists in 81 departments across the world in Carrasco and Ruiz-Castillo (2012); they are composed of 15.6 % of US scientists in computer sciences, engineering, and the sciences surveyed in Fox and Nikivincze (2021); top performers are also defined as the upper 10 % of Polish scientists surveyed in terms of article-equivalent productivity in Kwiek (2018a); they are 12 % of Italian STEMM scientists in terms of WoS their output in Abramo et al. (2009). The subsample of top performers examined can also be as low as 1 % of scientists studied as the "continuously publishing core" of the global scientific workforce as in Ioannidis et al. (2014); 0.1 % of scientists publishing in WoS as in O'Boyle and Aguinis (2012); and 0.1 % of most cited environmental scientists and ecologists based on Highly Cited Researchers' lists produced through WoS data as in Parker et al. (2010).

The research performance of star scientists has also been discussed from a gender perspective (Abramo et al., 2009). The stars were defined as scientists located in the top 10 % of the rankings of scientific performance, and they were examined using four indicators (output, fractional output, scientific strength based on the normalized impact factor of the relevant journal, and fractional scientific strength). The most intriguing finding was that men and women scientists do not differ much in terms of productivity for 90 % of the Italian scientific population, but they differ fundamentally among the upper 10 %, that is, the stars. As in our present research, in which we use a journal prestige-normalized approach, normalization of publications to journal prestige involved the transformation of absolute productivity values to productivity percentile ranks (Abramo et al., 2009), with the awareness of limitations of this methodological choice.

The various individual characteristics of top performers and characteristics of their institutions and the academic "climates" or "personalities" of their departments (e.g., Turner & Mairesse, 2005; Fox & Mohapatra, 2007) have been studied. The modest impact of their high productivity on the overall citation impact of their university departments has been explored (Piro et al., 2016). Removing the publications of the most productive scientists only marginally reduces the citation indexes because, although they have high productivity, their share in the total publication output of departments is always limited. Patterns of their academic collaboration (Abramo et al., 2011) and international mobility (Halevi et al., 2015) have also been examined. Top performers have been found to be overrepresented among the top earners across 10 European systems: Highly productive academics tend to receive higher salaries compared with their lower-performing colleagues at the same academic seniority levels (Kwiek, 2018b). Top performers may increase the productivity of those present in the organization (termed incumbents), and they may also increase the productivity of newly hired members (termed joiners); however, the effect on joiners—based on a difference-in-difference estimation—is reported to dominate the effect on incumbents (Agrawal et al., 2017).

Finally, high research productivity at the individual level has been found to be persistent over time (Turner & Mairesse, 2005; Kelchtermans & Veugelers, 2013). In a sample of French physicists, the vast majority of top performers (66 %) remained top performers within the 12 years studied (Turner & Mairesse, 2005), and about one-third of top performers in Italy (35 %) remained top performers in the 10 years studied (Abramo et al., 2017). Looking at 38 OECD countries, radical bottom-to-top (Jumpers-Up) and top-to-bottom (Droppers-Down) mobility between publishing productivity classes from a lifetime perspective was shown to be a very rare scholarly phenomenon when studied on a sample of 320,000 scientists with 25–50 years of publishing experience (Kwiek & Szymula, 2024b). A survey-based study ($N = 17{,}211$) has shown that the upper 10 % of scientists in terms of publishing productivity are responsible for about half of all research production across 11 European systems (Kwiek, 2016): what has been termed the 10/50 rule. Indeed, European top performers are a highly homogeneous group of academics whose high research productivity is driven by structurally similar factors.

In the Polish context, a recent longitudinal study of 2326 Polish full professors tracked over time for up to 40 years (Kwiek & Roszka, 2024) showed that previous top performance (e.g., being highly productive at earlier stages of academic careers) is a strong predictor increasing the odds of later top performance. In a different study based on a national survey of the academic profession ($N = 2593$) from the early 2010s, the upper 10 % of Polish scientists across all academic disciplines have been reported as being responsible for about 50 % of the total national research output (Kwiek, 2018a). The average research productivity distribution in Poland has also been reported to be highly skewed for its segment of top performers. The upper 10 % of academics are as internally stratified as the lower-performing 90 %, with a very small number of very high publishers. The right tail of the productivity distribution tends to behave exactly as the entire productivity distribution, demonstrating the scale invariance of Pareto curves (O'Boyle & Aguinis, 2012).

### 1.2. The focus of this research

The present research goes beyond cross-sectional self-reported survey data and provides a longitudinal view of publishing patterns in which all research-active Polish scientists in STEMM fields in the period of the three decades are examined. We had strong expectations about the scale of contribution to the total national research output by top performers based on previous survey-based research, but our current comprehensive longitudinal approach was not possible before the advent of large-scale datasets, that is, in our case, the complete Scopus raw dataset.

Our results are not valid for a single point in time (e.g., three years, as in most survey research); with panel data, long-term stability (or changes) can be examined. Time matters, and we use successive periods of time in our analysis. As Aguinis and O'Boyle noted in their study of star performers, "Stars are identified by their exceptional output over time and not just a single exceptional result" (2014: 315).

In almost all previous research (exceptions where 10–12 years were examined include Turner & Mairesse, 2005; Kelchtermans & Veugelers, 2013; and Abramo et al., 2017), the approaches to high research productivity are missing the time component. Cross-sectional studies, both national and international, were not complemented by longitudinal studies: Scientists comprising the classes of top performers (and the rest of the science profession) have not been tracked over time.

As a result, we ask a simple research question related to time: Are the patterns in the shares of national output that research top performers were producing in the past 30 years stable over time? Research top performers are discussed as the top 1 %, 5 %, and 10 % classes of scientists in terms of publishing productivity, respectively ranked within each STEMM discipline.

In the present research, we analyzed the role of research top performers in the national research output by gender, academic age (or publishing experience, as of 2021), academic discipline, institutional type (research-intensive institutions contrasted with the rest of them), and, most of all, changes over time for a period of 30 years. We used raw publishing and citation data from Scopus (both the data extracted for all Polish authors and institutions and the data made available to us through a multiyear agreement with Elsevier's International Center for the Study of Research ICSR Lab).

Research output at the individual level (limited to research articles and papers in conference proceedings) was computed for five 6-year study periods, three top performers classes, and 15 STEMM disciplines. We compared the output produced by two complementary subgroups, top performers, and the rest of the Polish research-active scientific population (e.g., by the upper 10% vs. the remaining 90 % of scientists, from the most inclusive to the most selective definition of top performers).

### 1.3. Research questions and hypotheses

The four research questions (Table 1) were based on selected findings from previous national and cross-national comparative studies on high research performance and data availability. Our hypotheses refer to the temporal stability in the share of output produced by top performers over the past 30 years (H1), disciplinary differentiation in this share (H2), unequal gender representation in top performance (H3), and the major predictors of membership in top-productivity classes (H4) based on logit generalized linear models with fixed effects.

## 2. Data, methodological approach, and sample

### 2.1. Data

We utilized the data on authors and papers extracted between November 2022 and January 2023 from the Scopus dataset, with complete data for the 1992–2021 period. We compiled a list of all Polish research-active institutions (343 in total, all institutional types, all sectors, no threshold of minimum publication number per year), and we compiled full lists of publication authors for each institution. The definition of scientists used in the current research is inclusive, including both publishing master's students (a very rare case in the Polish system) and publishing doctoral students. With the type of data we used, it was not possible to distinguish between academic scientists employed in institutions and students taught at these institutions as long as they were publishing in Scopus and listed under the same affiliation (Scopus institutional ID).

**Table 1**
Research questions, hypotheses, and summary of support.

| Research Questions | Hypotheses | Support |
|---|---|---|
| **RQ1.** Is the share of the output produced by top performers in the total national research output changing over time (1992–2021)? | **Temporal stability** **H1:** The role of top performers in producing the total national research output is relatively stable over time (for all classes of top performers). | Supported |
| **RQ2.** Is there disciplinary differentiation in the share of the output produced by top performers in the total national research output? | **Disciplinary differentiation** **H2:** The role of top performers differs substantially by discipline (for all classes of top performers). | Supported |
| **RQ3.** What is the relative presence of male and female scientists among top performers? | **Unequal gender representation** **H3:** Men scientists are overrepresented and women scientists underrepresented among top performers of all classes in the vast majority of STEMM disciplines. | Supported |
| **RQ4.** What are the predictors of membership in the classes of top performers? | **Model approach to top-productivity classes: logit generalized linear models with fixed effects** **H4:** Gender, academic age, and research collaboration patterns are the most important predictors of membership in the classes of top performers. | Partly supported |

Scientists assigned by Scopus to any Polish institution were defined as Polish scientists (in rare cases, when scientists had two or more affiliations, their publication counts were ascribed to these institutions). However, our approach uses individuals as a unit of analysis (as in our recent study of men and women leaving science in 38 OECD countries: Kwiek & Szymula, 2024a); hence, individual productivity is not changed by scientists' having two or more affiliations (or different affiliations, national and international, changing over time), as rare as these cases may be. Our fine-grained approach allowed us to assign institutional affiliation to scientists within the five 6-year periods, here based on all publication bylines from each period. Only in regression analysis do we use the institutional research intensity variable, and in these cases, the 2021 Scopus affiliation is used as a proxy.

Information about the authors included their full names, the disciplines in which they most frequently (the modal value) published their articles and papers in conference proceedings, information about their affiliations, and their unique Scopus ID. We used complete bibliometric metadata about the authors, along with their Scopus ID. In addition, using the database from the ICSR Lab, the year of each scientist's first publication was obtained, which we used to define academic age (or publishing experience) at the time of publishing any article and at the end of each six-year period.

As part of data cleaning, at the author level, the following cases were removed from the analysis: 6885 cases (Polish scientists only) for which gender could not be defined (e.g., initials only, incomplete first names, non-Polish first names, wrongly formatted specifically Polish characters in first names, last names used as first names); 14,431 cases with an unknown year of the first publication (e.g., their first article was published prior to 1992; these authors were present in the authors' database but not in the publication database for 1992–2021); and 49,939 cases for which the dominant discipline could not be determined (e.g., there were no cited references in papers in Scopus database). As part of data cleaning at the level of publications, papers with >100 authors (mainly in PHYS)—in which true authorship versus other types of contribution is hard to determine, not to mention mega-author titles—were arbitrarily removed (see Dotson, 2024 on this limitation).

In the next step of data preprocessing, only journal articles and papers in conference proceedings were selected. Finally, the non-STEMM disciplines and disciplines with small numbers of authors in our sample (DEC, HEALTH, IMMU, NURS, VET, BUS, ECON, HUM, PSYCH, SOC, DENT, and MULTI) were removed from the analysis. In some cases, a meaningful binary division of our sample into the upper 10 % and rest in terms of productivity was impossible: The productivity distribution was too flat, with no tails on the right. Only at this point of data preprocessing was the article dataset linked deterministically (using individual Scopus author IDs as the key) to the authors' database. Finally, the dataset contained 152,043 Polish authors (and their 587,558 articles from the 1992–2021 period).

### *2.2. Methodological approach*

Our methodological approach is as follows:

(1) *Longitudinal*: we track scientists over time up to five 6-year periods of time, and maximally between 1992 and 2021.
(2) *Relative*: we compare scientists using their productivity ranks and allocating them to productivity percentiles, from the top to the bottom percentiles, rather than using actual publication numbers.
(3) *Classificatory*: we classify all scientists based on their productivity ranks and compare the output of the three top performers' classes with the output by the rest of the scientists.
(4) *Prestige-normalized*: we recognize the highly stratified nature of the current journal-based global science system, highlighting the average differences in scholarly effort and impact on the science community between publishing in generally low-impact journals and generally high-impact journals.

Our unit of analysis was the individual scientist with a separate publishing history in any or all of the five 6-year periods. We divided the study period (1992–2021) into five 6-year periods. Separately within 15 STEMM disciplines, we ranked all internationally visible scientists (i.e., scientists publishing research articles in Scopus-indexed journals) from the most productive to the least productive, and we created the three classes of research top performers (the upper 1 %, 5 %, and 10 %) in each study period.

For every scientist, we used a dominant academic discipline within a period. Dominant disciplines were computed based on the modal value (or randomly selected value from among the dominants if there was more than one) from all disciplines (All Science Journal Classification used in Scopus, ASJC) assigned to the journals of all cited references in all papers in scientists' publications in a given six-year period.

In other words, using the ICSR Lab database, all cited references from all publications from a period were linked to ASJC disciplines, and their modal value for a period was selected. As a result, scientists could be assigned to the same or different STEMM disciplines as their careers developed—which is a more fine-grained approach to discipline assignment than using all the cited references from all publications (lifetime) to select a single dominant discipline for the whole career.

The ASJC classification offers three levels, from the most detailed level of 333 unique fields of research to 27 subject-level fields to four fields (physical sciences, life sciences, health sciences, social sciences). We used the 2-digit (27 subject-level fields) Scopus classification codes rather than the 4-digit (333 fields of research) Scopus codes because the number of observations at the most detailed level was far too limited for any meaningful analysis, especially for the oldest period (1992–1997) and for the analyses of the representation of women among top performers. We could easily identify top performers at the most detailed level, but the cut-off points between top performers (especially for women in the top 1 % class) and the rest would be hard to handle. There may be variability within so defined disciplines—and we list this methodological choice among our limitations.

The gender of scientists (binary: male, female) was defined using *genderize.io*, a gender-determination software, with the probability threshold set at 0.85. Using the ICSR Lab database, we obtained the year of each scientist's first publication, allowing us to determine their academic age at any point in time (or for any publication) and in 2021 (we followed the same procedure in our study of the changing demographics of the global science workforce in Kwiek and Szymula, 2023).

Institutional affiliation (binary: research-intensive institutions, rest of institutions) was defined as the dominating affiliation in a given six-year period based on publications' bylines from this period. We have defined research-intensive institutions as those following an IDUB national excellence program for 2020–2026 in which 10 Polish universities and technical universities were defined as research excellent, with additional research funding; however, a small sector of the institutes of the Polish Academy of Sciences (PAS) in this binary distinction belongs to the rest of the institutions, which we list among the limitations.

Our approach to productivity is journal prestige-normalized (and uses a full-counting method in which equal credit goes to each coauthor in multiauthored publications), highlighting the difference in average scholarly efforts and average impact on the global scholarly community through citations between articles published in generally unselective and low-prestige journals and in generally highly selective and high-prestige journals (see a study of duplicate articles and their citations in Larivière & Gingras, 2010).

Linking individual publications to journals and their global percentile ranks (in the range of 0–99) was feasible; however, linking individual publications to their per-year citations over time (or to their field-weighted citation impact for a specific time window, e.g., four years after the publication) was not feasible with our dataset. Although working with article-based impact metrics would be much more preferable, the limitations of our dataset forced us to use only journal-based impact metrics, with all their limitations discussed in the literature (see, e.g., Weingart, 2004). An improvement on the present study's results would be to implement productivity measures based on the citation impact directly achieved by each individual publication (as suggested but not implemented in Carrasco & Ruiz-Castillo, 2012).

Still, in the specific context of the subsequent Polish national research assessment exercises of the past three decades, the approach using journal prestige normalization (separately within Scopus disciplines) works better than those approaches without any normalization to citations, be they full counting or fractional counting.

In Poland, the institutional, disciplinary, and individual assessments of research outputs have been routinely conducted through so-termed publication points assigned to papers through ministerial journal lists (mostly closely related to Scopus and WoS impact data, from top journals being assigned 200 or 140 points to the least prestigious journals with 20 points). In the current research, we follow the fundamental concept of Polish research assessment exercises: "Academic journals are not equal." In Poland, academic careers have been closely linked to papers in journals with high numbers of points, with journal types in one's publishing portfolios having powerful implications for higher chances in research grant acquisition and faster academic promotions in the university system.

We have also computed the data and prepared full analyses, including regression models, for three other productivity types—prestige-normalized, fractional counting; standard non-normalized, full counting; and standard non-normalized, fractional counting—with very similar patterns and regression results; however, space limitations precluded the presentation and comparative analysis of the four productivity types.

The journal prestige normalization used is citation based, and journal percentile ranks are determined annually in Scopus, here based on all citations received by all documents published by a journal relative to their number (in the previous four years). As a result, articles in journals located higher in the Scopus CiteScore percentile ranks ($N$ = 43,092 journals, 2024) were given more weight in computing productivity than those in journals located lower (ranks range: 0–99). In a non-normalized approach to productivity (full-counting method), articles published in all journals would receive a value of 1, whereas, in our prestige-normalized approach (full-counting method), articles in journals with a percentile rank of 90 received a value of 0.90; articles published in journals (and papers in conference proceedings) with percentile ranks of 10 and less—including articles published in journals with the lowest percentile rank (but still indexed in Scopus)—received a value of 0.1. In the Polish system, journals indexed in Scopus (binary: yes/no) with the lowest CiteScore percentile rank (0 %) are still valued higher in research assessment exercises than those publications in journals not indexed in Scopus. Our dataset includes only Scopus-indexed journals, so publications in journals in the lowest percentile (0 %, CiteScore mostly 0.0 and 0.1) are included.

We study the productivity of individual scientists in relation to the productivity of other individual scientists within the same disciplines, and we use classes of scientists rather than publication numbers. As a result, we avoid issues related to increasing numbers of journals indexed in Scopus over time and increasing participation of Polish scientists in them over time.

### *2.3. Sample*

Our sampling approach to top performers was relatively simple: We studied scientists (and top performers, their upper layers of 1 %, 5 %, and 10 %) separately within the five 6-year periods and separately within disciplines. Depending on their academic age or the length of their publishing experience, some scientists (and some top performers) appeared in a single period and others in several or all periods. For scientists (and top performers) to be included in computations for a given period, they had to have at least a single article published in Scopus in this period.

The six-year segments were selected to ensure that the number of top performers per discipline and by gender in each segment met the minimum requirements (this was especially important to examine the role of male and female top performers in such male-dominated disciplines as COMP, ENG, MATH, and PHYS). However, we have also examined four-year and five-year time segmentation to see whether the results are sensitive to different time segmentations (see Electronic Supplementary Material).

The subsets from which top performers were selected increase by period: from 23,166 scientists in the first period of 1992–1997 to 93,092 in the last period of 2016–2021 (Table 2). The upper 10 %, 5 %, and 1 % classes in terms of productivity increase accordingly: for example, the upper 10 % class increases from 2400 in the first period to 9337 in the last period (Table 3).

The subsets of scientists from which top performers were selected may be overlapping in the five periods, but it does not have to be so: Some scientists published in the first two periods (1992–1997 and 1998–2003) and others in the second and third period (1998–2003 and 2004–2009); finally, the youngest scientists published in the last period only (2016–2021). Our unique authors (152,043 scientists) appear in any or all periods; hence, the number of scientists is all periods (Table 2) does not add up to 152,043 but is much higher.

We did not work with the 1992 cohort (only scientists who started publishing in 1992): We worked with all scientists publishing in any or all periods between 1992 and 2021. The same scientists could be top performers in one period and belong to the rest of scientists in other periods. The distribution of top performers by top performers class, period, gender, institutional type, and academic age group is shown in Supplementary Table 1.

About a quarter of all scientists in the five periods studied come from research-intensive universities and about three-quarters from other universities (i.e., the rest). The percentage of top performers coming from research-intensive institutions remains almost unchanged over the analyzed period (and remains in the range of 25–32 %; Supplementary Table 1). The percentage is unexpectedly low, is stable over time, and does not increase as we move up the high-performance class from the upper 10 % to the upper 1 %. Scientists from research-intensive universities are not overrepresented, as could be expected: their shares in the subpopulation of top performers reflect their shares in the total population of scientists.

In general, the share of women among top performers of all three classes increases with each subsequent period (e.g., from about a quarter in the first to almost 40 % in the last period in the case of the top 10 %). The share of women among the top 1 % almost doubles, reaching a quarter in the last period. However, for all periods, the share of women decreases with each subsequent top performance class: It is always the highest among the top 10 % and lowest among the top 1 %.

In the present research, our focus is on 15 STEMM disciplines: AGRI, agricultural and biological sciences; BIO, biochemistry, genetics, and molecular biology; CHEMENG, chemical engineering; CHEM, chemistry; COMP, computer science; EARTH, earth and planetary sciences; ENER, energy; ENG, engineering; ENVIR, environmental science; MATER, materials science; MATH, mathematics; MED, medical sciences; NEURO, neuroscience; PHARM, pharmacology, toxicology, and pharmaceutics; and PHYS, physics and astronomy.

## 3. Results

### 3.1. Empirical patterns: contribution of top performers to the total national research output over 30 years

Top performers produce a significant—and almost unchanging over time—proportion of Poland's total national research output in the five 6-year periods into which we have divided the past 30 years. For the top 10 %, the proportion in the latest study period is 45.5 % for all disciplines combined, and for the top 1 %, the proportion is 10.8 % for all disciplines combined. Overall, the proportions for all periods are in the range of 43.8–46.6 % for the top 10 % and 10.1–11.6 % for the top 1 %, with interesting disciplinary differentiations (Table 4).

**Table 2**
The distribution of the total sample from which top performers were selected by six-year period, gender, institutional type, academic age (publishing experience) group in 2021, and STEMM discipline (frequencies).

| Periods | | 1992–1997 | 1998–2003 | 2004–2009 | 2010–2015 | 2016–2021 |
|---|---|---|---|---|---|---|
| **Total** | | **23,166** | **36,366** | **54,346** | **76,310** | **93,092** |
| Gender | Female | 8480 | 15,081 | 25,104 | 36,743 | 47,155 |
| | Male | 14,686 | 21,285 | 29,242 | 39,567 | 45,937 |
| Research intensity | Research-intensive | 6338 | 9889 | 14,216 | 20,306 | 25,232 |
| | Rest | 16,828 | 26,477 | 40,130 | 56,004 | 67,860 |
| Academic age groups | 0–9 years | 14,529 | 24,170 | 35,278 | 47,396 | 54,040 |
| | 10–19 years | 5297 | 6382 | 10,684 | 17,728 | 22,489 |
| | 20–29 years | 2853 | 4350 | 5132 | 6098 | 10,256 |
| | 30 and more years | 487 | 1464 | 3252 | 5088 | 6307 |
| Discipline | AGRI | 1715 | 2916 | 4789 | 7649 | 9336 |
| | BIO | 2524 | 3766 | 5199 | 7280 | 8922 |
| | CHEM | 3318 | 4818 | 6159 | 7731 | 8221 |
| | CHEMENG | 220 | 375 | 491 | 535 | 570 |
| | COMP | 188 | 338 | 762 | 1620 | 1831 |
| | EARTH | 951 | 1594 | 1971 | 2712 | 3367 |
| | ENER | 50 | 109 | 223 | 559 | 1036 |
| | ENG | 1065 | 1788 | 3370 | 5681 | 7952 |
| | ENVIR | 421 | 767 | 1456 | 2636 | 3928 |
| | MATER | 891 | 1331 | 2207 | 3730 | 5019 |
| | MATH | 898 | 1248 | 1705 | 2204 | 2244 |
| | MED | 6900 | 12,224 | 20,143 | 26,929 | 33,167 |
| | NEURO | 230 | 366 | 455 | 652 | 892 |
| | PHARM | 375 | 410 | 420 | 509 | 535 |
| | PHYS | 3420 | 4316 | 4996 | 5883 | 6072 |

**Table 3**
The distribution of the subsample of top performers (the top 10 %, 5 %, and 1 % scientists (in terms of productivity) by six-year period (% and frequencies).

| | Top 10 % | The top 5 % | The top 1 % |
|---|---|---|---|
| 1992–1997 | 2400 | 1202 | 241 |
| 1998–2003 | 3702 | 1846 | 374 |
| 2004–2009 | 5463 | 2726 | 550 |
| 2010–2015 | 7682 | 3841 | 770 |
| 2016–2021 | 9337 | 4666 | 934 |

Our results show that, from a long-term perspective, the structure of the Polish national output in terms of publications by the class of top performers versus the rest of the scientists remains surprisingly stable. This stability may be related to the stability in the distribution of publishing productivity in the population of all scientists over time. In the simplest terms, on average, the top 1 % account for 10 % of the national output, and the top 10 % account for almost half of the national output. These proportions hardly change over time, meaning that the processes of increasing the concentration of publications in the hands of the most productive scientists are not observed.

However, significant disciplinary variations in each of the analyzed periods are observed. In each period, the highest proportion of national output for the top 10 % was observed for PHARM (as much as 62.1 % in the first period and 58.1 % in the last period). For the largest discipline, MED, the proportion approached 50 % for the last three periods. For the top 1 %, the largest proportion was also observed for PHARM in the first, second, and last periods; and for MED, the proportion ranged from 10 % to 13.0 % over the entire study period.

**Table 4**
Publications by the top 1 % and 10 % of scientists as a percentage of all publications (the total national output), by STEMM discipline and period (in %).

| *The top 1 %* | | | | | |
|---|---|---|---|---|---|
| | 1992–1997 | 1998–2003 | 2004–2009 | 2010–2015 | 2016–2021 |
| AGRI | 9.0 | 10.0 | 9.4 | 10.0 | 9.3 |
| BIO | 9.6 | 9.3 | 9.4 | 9.6 | 8.9 |
| CHEM | 9.7 | 10.8 | 11.9 | 12.1 | 11.3 |
| CHEMENG | 15.5 | 13.8 | 11.2 | 11.8 | 9.7 |
| COMP | 8.4 | 9.8 | 11.9 | 12.5 | 12.0 |
| EARTH | 9.2 | 10.5 | 10.9 | 10.1 | 9.2 |
| ENER | 7.6 | 9.4 | 8.4 | 13.8 | 11.0 |
| ENG | 9.2 | 11.2 | 12.0 | 11.2 | 11.1 |
| ENVI | 11.1 | 13.8 | 13.4 | 10.7 | 10.1 |
| MATER | 9.5 | 10.0 | 10.6 | 9.8 | 10.3 |
| MATH | 7.9 | 7.9 | 8.2 | 8.7 | 8.9 |
| MED | 10.7 | 10.8 | 12.9 | 12.4 | 11.9 |
| NEURO | 8.4 | 7.4 | 9.6 | 9.1 | 8.7 |
| PHARM | 19.2 | 15.6 | 9.4 | 12.0 | 13.2 |
| PHYS | 10.5 | 11.9 | 11.8 | 12.9 | 11.6 |
| **TOTAL** | **10.1** | **10.7** | **11.6** | **11.5** | **10.8** |
| ***The top 10 %*** | | | | | |
| | **1992–1997** | **1998–2003** | **2004–2009** | **2010–2015** | **2016–2021** |
| AGRI | 39.0 | 42.5 | 42.2 | 43.1 | 42.7 |
| BIO | 41.7 | 41.8 | 41.7 | 43.0 | 41.8 |
| CHEM | 45.0 | 46.5 | 47.7 | 47.3 | 45.6 |
| CHEMENG | 46.5 | 47.3 | 46.1 | 44.9 | 45.6 |
| COMP | 39.5 | 45.9 | 44.6 | 45.9 | 45.0 |
| EARTH | 43.8 | 45.0 | 45.1 | 43.9 | 41.3 |
| ENER | 33.2 | 35.2 | 38.0 | 46.7 | 43.6 |
| ENG | 41.4 | 44.1 | 44.2 | 44.5 | 42.3 |
| ENVI | 44.1 | 46.1 | 46.2 | 43.6 | 43.5 |
| MATER | 42.5 | 44.6 | 44.2 | 43.9 | 45.8 |
| MATH | 35.6 | 37.8 | 40.0 | 41.6 | 39.3 |
| MED | 45.7 | 46.6 | 49.8 | 49.1 | 49.0 |
| NEURO | 41.7 | 42.2 | 47.2 | 41.3 | 39.2 |
| PHARM | 62.1 | 58.5 | 52.3 | 52.7 | 58.1 |
| PHYS | 44.1 | 46.8 | 45.8 | 47.2 | 45.1 |
| **TOTAL** | **43.8** | **45.4** | **46.6** | **46.4** | **45.5** |

At the same time, top performers do not form homogeneous classes. They are represented by men and women; they are affiliated with more research-intensive (Top) and less research-intensive (Rest) institutions; and they belong to different academic age groups (or groups of publishing experience calculated using the time elapsed since their first publication) (Table 5). Using academic age as a proxy for biological age (with correlations analyzed in detail for Poland Kwiek & Roszka, 2022), we refer to them as beginners, early-career, midcareer, and late-career scientists (academic age groups 0–9, 10–19, 20–29, and 30 and more years of publishing experience, respectively).

Our analyses show substantial gender differences among the top 10 % (Table 5). Although women in the top 10 % are responsible for about one-third of all publications produced by women (i.e., the rest of women are responsible for about two-thirds of all publications produced by women), with the share increasing over time, the concentration of research is substantially higher among the male top 10 %. The male top 10 % have been consistently responsible for more than 50 % of all publications produced by men. The differences between the top 10 % affiliated with research intensive versus with all other institutions are marginal across the 30 years studied. For all periods among midcareer and late-career scientists, the share of publications produced by top performers exceeds 60 %. Similar patterns emerge for the top 1 % and 5 %, which are not reported here because of space limitations.

The proportions of publications produced by the top 10 %, 5 %, and 1 % of the total national output by discipline clearly show a high concentration of publishing productivity that is stable over time. A small group of scientists has generated a significant proportion of work in their disciplines at similar levels across all analyzed periods. In some disciplines, the contribution of top performers is high across all periods—and in others, it is clearly lower. For example, PHARM clearly dominates in terms of concentration levels, which remain in the 52–62 % range in the period studied. In contrast, MATH is characterized by a radically lower concentration of output (at the level of 36—42 %), indicating a more even distribution of publications: Mathematicians tend to contribute more equally to the total national output in mathematics. Small fluctuations can be observed for disciplines between successive periods, but these fluctuations do not translate into clear long-term trends.

### 3.2. *The relative presence index (RPI) among top performers for men and women*

The classes of the most productive scientists consist of male and female scientists, whose proportional contributions to the total national output vary over time and differ between the selected STEMM disciplines. In seemingly homogeneous classes of top performers, gender differences in proportional contributions to the total national output can be observed.

In this section, we are interested in the *relative* presence of female top performers among top performers. We normalize the number of women among the top performers in a given period and discipline to the number of all women in the population of scientists in that period and discipline.

A higher proportion of men in a discipline for one period should be reflected by a higher proportion of men among the top performers in that discipline in that period. This relative approach to the presence of women among top performers allows us to show their role from a comparative perspective of disciplines—and from a comparative perspective of time—over the five 6-year periods (avoiding a composition effect, whereby apparent gender differences can be ascribed to the composition of the group examined; see Nygaard et al., 2022; see also Kwiek & Roszka, 2021, on gender homophily in research collaboration patterns in Poland).

As a result, we have constructed what we term the Relative Presence Index (RPI) among top performers. For women in the class of top performers, an RPI of 1 means the same relative presence of men and women among top performers. The RPI for women can show the level of their overrepresentation (value above 1) or underrepresentation (value below 1) among top performers. Although the numbers (in a nominal approach) may confirm the simple fact that male top performers in a given discipline outnumber female top performers, the RPI for women provides a more adequate and intuitively understandable measure of their presence among top performers.

The RPI for men is constructed as the number of male top performers in relation to the total number of all men (in a given domain) divided by the number of female top performers in relation to the total number of all women (in a given domain), as follows:

$$RPI = (n(TP_men) / n(all_men))/(n(TP_women) / n(all_women))$$

The RPI can vary over time and between disciplines. In technical terms, the RPI is the quotient of two quotients: male top performers/all men divided by female top performers/all women (see Abramo et al., 2009 on “star scientists”). The relationship between the RPI for men and RPI for women is harmonic (rather than linear). If the women in a certain domain show an RPI of 0.33, then men show an RPI of 3.03 (1/0.33) in this domain; if the women show an RPI of 0.5, then men show an RPI of 2 (1/0.5).

**Table 5**
Publications by the top 10 % as a percentage of all publications (the total national output), by gender, institutional type, and career age group, with all STEMM disciplines combined (in %).

| Periods | Total | Gender | | Institutional research intensity | | Career age groups (length of publishing experience) | | | |
|---|---|---|---|---|---|---|---|---|---|
| | | Female | Male | Research intensive | Rest | Beginners (0–9 years) | Early-career scientists (10–19 years) | Midcareer scientists (20–29 years) | Late-career scientists (30 and more years) |
| 1992–1997 | 43.8 | 34.7 | 47.6 | 41.7 | 44.7 | 26.3 | 50.5 | 62.0 | 68.0 |
| 1998–2003 | 45.4 | 36.0 | 50.1 | 45.8 | 45.2 | 25.3 | 55.5 | 61.6 | 66.9 |
| 2004–2009 | 46.6 | 37.3 | 52.3 | 47.4 | 46.3 | 24.1 | 56.2 | 62.6 | 69.6 |
| 2010–2015 | 46.4 | 37.8 | 52.4 | 47.7 | 45.9 | 22.6 | 54.4 | 65.0 | 67.8 |
| 2016–2021 | 45.5 | 38.4 | 51.0 | 46.0 | 45.2 | 23.0 | 52.5 | 61.6 | 61.7 |

How should the indexes of 1.29 for men and 0.77 for women for AGRI for the top 10 % (in the period 1992–1997, Table 6) be interpreted? The indexes indicate that the representation of men in the top 10 % is 29 % higher than the representation of women in the top 10 % in this discipline for this period. Thus, men are overrepresented in the top 10 % compared with women. Conversely, the representation of women in the top 10 % is only 61 % of the representation of men in the top 10 % in this discipline during this period. Therefore, women are underrepresented in the top 10 % (at the same time, the relationship between the indexes is harmonic: 1/1.29 = 0.77, with rounding). For the sake of showing two perspectives (generally men being overrepresented and women being underrepresented), we retain the results for both men and women in Table 6; harmonic relationships are better shown for the two subsamples.

The relative overrepresentation of men among the top 10 % and top 1 % (Table 6) as well as among the top 5 % (Supplementary Table 2) can be examined through several dimensions: by period, that is, changing over time; by academic age group, that is, changing by seniority; and by discipline.

The general patterns show that men are overrepresented among each top performers class, and this overrepresentation increases when moving up the high-performance scale. In the final period studied (2016–2021), the RPI for the male top 10 % is 1.59 and 2.97 for the top male 1 % (all disciplines combined: Total in Table 6). In terms of academic age groups, the index generally decreases for scientists with long academic experience, and for the top 10 %, it approaches 1, though never reaching it. The RPI for men increases when moving up the high-performance scale, with a very large overrepresentation of men found among the top 1 %. In the five periods studied, for the top 10 % and top 5 % (but not for the top 1 %), the RPI remains at similar levels. Finally, in terms of discipline, the RPI is highly diversified. However, no simple relationship between the RPI and composition of the discipline by gender seems to apply. The RPI for women is inversely proportional to the RPI for men, and its interpretation is the inverse of the interpretation of the index for men.

## 4. Multidimensional logistic regression approach

### *4.1. Logit generalized linear models with fixed effects*

To test the determinants of membership in the classes of the most productive scientists, we have developed an econometric model. Because of the binary nature of the dependent variable, we used a logistic regression model with fixed effects (Allison, 2000). The model was cross-sectional and dynamic (Fernández-Val & Weidner, 2016; Hinz et al., 2020; Stammann, 2018). The fixed effects in the model are periods and STEMM disciplines. Five 6-year publishing period in 1992–2021 and 15 disciplines have been used. Disciplines in periods were the dominant disciplines (the mode) in the author's full lists of cited references in all their publications in a given period, according to Scopus ASJC journal classification based on 27 subject-level fields, here with 2-digit codes used. The independent and fixed effects variables are described in Table 7.

The model is cross-sectional because the population is made up of individual scientists (the unit of analysis being a single scientist described by a vector of characteristics), and it is dynamic because there are five disjointed six-year periods. The inclusion of the time variable in the model as an independent variable would disturb the assumption of the independence of observations because individual scientists can publish in more than one period (and some publish in all periods). At the same time, the periods differed in their social and economic characteristics (conditioning the functioning of the public science system). An additional advantage of periods treated as a fixed effect in the model was that publishing productivity could be compared between periods without the interference of external changes. The choice of disciplines as a second fixed effect was directed by internal variations in publishing patterns within disciplines and the possibility of making cross-disciplinary comparisons without distortions being associated with the influence of discipline-specific factors on overall productivity. R software version 4.3.0, RStudio version 2023.06.0 Build 421, and the "alpaca" package version 0.3.4 were used for the computations. The R package "alpaca" is a specialized tool for modeling complex models with fixed effects. Specifically, the model was estimated using the *feglm* (Fixed Effects Generalized Linear Model) function, which allows for the inclusion of all variable categories in the model, differing from the traditional approach where one category is typically treated as a reference. The "alpaca" package uses a specific estimation technique known as absorption of multicollinearity. This method eliminates the issue of multicollinearity among dummy variables, even when all categories of categorical variables are included in the model. Instead of excluding one category as a reference, as is common in traditional approaches, the *feglm* function absorbs the redundant information, ensuring that the estimation remains correct and stable.

In models constructed with the *feglm* function, full coding of variables is used. This means that all categories of categorical variables are represented separately in the model, allowing for the estimation of the effect of each category without the need to refer to a reference category. This approach provides a more comprehensive view of the impact of each category on the dependent variable. In this model, all categories are treated as separate variables, allowing for the direct estimation of the effect of each category on the dependent variable. As a result, there is no need to compare results with a reference category since all effects are presented relative to the average level of the effect.

Each fixed effect in the model can be interpreted as the average impact of a given category on the dependent variable, taking into account all other effects in the model. This provides full information on the influence of each category on the dependent variable. In summary, the use of the "alpaca" package and the *feglm* function enabled a more comprehensive approach to modeling with fixed effects, incorporating all variable categories and providing a fuller interpretation of the results.

**Table 6**
The Relative Presence Index (RPI) for men (left panels) and for women (right panels) top performers, the top 10 % and top 1 % (in terms of productivity) by six-year period, academic age group, and STEMM discipline.

| Variable | Category | 1992–1997 Men | 1998–2003 | 2004–2009 | 2010–2015 | 2016–2021 | 1992–1997 Women | 1998–2003 | 2004–2009 | 2010–2015 | 2016–2021 |
|---|---|---|---|---|---|---|---|---|---|---|---|
| | | *The top 10 %* | | | | | | | | | |
| **Academic age** | 0–9 years | 1.63 | 1.40 | 1.50 | 1.48 | 1.73 | 0.61 | 0.71 | 0.66 | 0.68 | 0.58 |
| | 10–19 years | 1.29 | 1.31 | 1.31 | 1.25 | 1.19 | 0.77 | 0.76 | 0.76 | 0.80 | 0.84 |
| | 20–29 years | 1.37 | 1.20 | 1.18 | 1.16 | 1.08 | 0.73 | 0.83 | 0.85 | 0.86 | 0.92 |
| | 30 & more years | 1.03 | 1.28 | 1.16 | 1.15 | 1.06 | 0.97 | 0.78 | 0.86 | 0.87 | 0.94 |
| **Discipline** | AGRI | 1.29 | 1.42 | 1.56 | 1.46 | 1.52 | 0.77 | 0.70 | 0.64 | 0.69 | 0.66 |
| | BIO | 2.19 | 2.33 | 2.51 | 2.08 | 1.91 | 0.46 | 0.43 | 0.40 | 0.48 | 0.52 |
| | CHEM | 1.97 | 1.96 | 1.98 | 1.98 | 1.83 | 0.51 | 0.51 | 0.50 | 0.51 | 0.55 |
| | CHEMENG | 1.91 | 1.75 | 1.86 | 2.17 | 2.35 | 0.52 | 0.57 | 0.54 | 0.46 | 0.42 |
| | COMP | 1.55 | 0.63 | 0.88 | 0.94 | 2.00 | 0.65 | 1.59 | 1.14 | 1.06 | 0.50 |
| | EARTH | 2.05 | 1.86 | 1.96 | 1.89 | 1.35 | 0.49 | 0.54 | 0.51 | 0.53 | 0.74 |
| | ENER | 0.44 | 1.19 | 1.69 | 1.07 | 1.76 | 2.28 | 0.84 | 0.59 | 0.94 | 0.57 |
| | ENG | 3.03 | 1.95 | 1.86 | 1.97 | 1.74 | 0.33 | 0.51 | 0.54 | 0.51 | 0.58 |
| | ENVIR | 1.01 | 0.99 | 1.04 | 1.29 | 1.11 | 0.99 | 1.01 | 0.96 | 0.78 | 0.90 |
| | MATER | 1.37 | 1.51 | 1.53 | 1.41 | 1.43 | 0.73 | 0.66 | 0.65 | 0.71 | 0.70 |
| | MATH | 3.35 | 18.14 | 2.71 | 1.88 | 1.14 | 0.30 | 0.06 | 0.37 | 0.53 | 0.88 |
| | MED | 1.34 | 1.46 | 1.75 | 1.84 | 1.79 | 0.75 | 0.68 | 0.57 | 0.54 | 0.56 |
| | NEURO | 1.21 | 1.46 | 1.22 | 1.37 | 1.00 | 0.82 | 0.68 | 0.82 | 0.73 | 1.00 |
| | PHARM | 0.98 | 0.74 | 0.89 | 1.26 | 1.61 | 1.02 | 1.35 | 1.13 | 0.80 | 0.62 |
| | PHYS | 1.90 | 1.66 | 1.82 | 1.57 | 1.60 | 0.53 | 0.60 | 0.55 | 0.64 | 0.63 |
| | **Total** | **1.56** | **1.60** | **1.71** | **1.66** | **1.59** | **0.64** | **0.62** | **0.58** | **0.60** | **0.63** |
| | | ***The top 1 %*** | | | | | | | | | |
| **Academic age** | 0–9 years | 2.94 | 2.82 | 3.54 | 2.54 | 3.75 | 0.34 | 0.35 | 0.28 | 0.39 | 0.27 |
| | 10–19 years | 3.55 | 1.79 | 2.14 | 1.96 | 2.16 | 0.28 | 0.56 | 0.47 | 0.51 | 0.46 |
| | 20–29 years | 3.16 | 2.26 | 1.57 | 1.66 | 1.80 | 0.32 | 0.44 | 0.64 | 0.60 | 0.55 |
| | 30 & more years | 2.45 | 1.98 | 1.57 | 1.70 | 1.72 | 0.41 | 0.51 | 0.64 | 0.59 | 0.58 |
| **Discipline** | AGRI | 4.28 | 2.57 | 2.75 | 3.63 | 2.73 | 0.23 | 0.39 | 0.36 | 0.28 | 0.37 |
| | BIO | 4.63 | 6.32 | 8.76 | 5.42 | 3.84 | 0.22 | 0.16 | 0.11 | 0.18 | 0.26 |
| | CHEM | 9.01 | 3.60 | 2.74 | 2.64 | 2.61 | 0.11 | 0.28 | 0.36 | 0.38 | 0.38 |
| | CHEMENG | - | - | - | - | - | – | – | – | – | – |
| | COMP | - | - | 1.27 | 0.79 | 3.71 | – | – | 0.79 | 1.26 | 0.27 |
| | EARTH | - | 3.32 | 2.17 | 1.96 | 2.09 | – | 0.30 | 0.46 | 0.51 | 0.48 |
| | ENER | - | 0.24 | - | - | 2.09 | – | 4.19 | – | – | 0.48 |
| | ENG | - | - | 5.90 | 5.04 | 2.59 | – | – | 0.17 | 0.20 | 0.39 |
| | ENVIR | - | 2.98 | 2.17 | 2.09 | 1.90 | – | 0.34 | 0.46 | 0.48 | 0.53 |
| | MATER | 0.51 | 3.11 | 2.46 | 1.95 | 2.72 | 1.95 | 0.32 | 0.41 | 0.51 | 0.37 |
| | MATH | - | - | 5.79 | 1.11 | 3.25 | – | – | 0.17 | 0.90 | 0.31 |
| | MED | 2.65 | 2.60 | 2.84 | 3.35 | 4.05 | 0.38 | 0.38 | 0.35 | 0.30 | 0.25 |
| | NEURO | 1.32 | 3.08 | 2.69 | 1.27 | 1.55 | 0.76 | 0.32 | 0.37 | 0.79 | 0.65 |
| | PHARM | 4.11 | 0.36 | 1.71 | 0.51 | 1.61 | 0.24 | 2.80 | 0.58 | 1.95 | 0.62 |
| | PHYS | 8.89 | 2.20 | 2.77 | 3.99 | 3.56 | 0.11 | 0.45 | 0.36 | 0.25 | 0.28 |
| | **Total** | **3.64** | **2.92** | **2.95** | **2.85** | **2.97** | **0.27** | **0.34** | **0.34** | **0.35** | **0.34** |

Note: "-" = no representatives of women scientists in this category.

**Table 7**
Description of independent and fixed effects variables.

| Variable | Definition | Type of variable |
|---|---|---|
| Academic age in period | The average difference between the year ending the period (1997, 2003, 2009, etc.) and the year of the author's first publication | Quantitative continuous |
| Average team size in period | The arithmetic average of the size of the teams (the number of coauthors plus 1 for every article) in a given period | Quantitative continuous |
| General collaboration rate in period | The percentage of articles written in collaboration: the ratio of the number of articles with one or more coauthors divided by the total number of articles in a given period (range 0–100 %) | Quantitative continuous |
| International collaboration rate in period | The percentage of articles written in international collaboration: the ratio of the number of articles written in collaboration with an author with a foreign country affiliation divided by the total number of articles in a given period (range 0–100 %) | Quantitative continuous |
| Gender | Author's gender | Qualitative, binary (male, female) |
| Institutional research intensity | Affiliation with one of the 10 institutions recognized as research-intensive institutions in the national IDUB research excellence program (2019–2026) | Qualitative, binary (Research-intensive, rest) |
| Discipline | Dominant discipline (the mode) in the author's full lists of cited references in all publications in a given period, according to Scopus ASJC journal classification, 27 subject-level fields, 2-digit coded used | Qualitative, 15-variant |
| Period | Six-year publishing period, five periods for 1992–2021 | Qualitative, 5-variant |

### *4.2. Independent variables*

The selection of variables (Table 7) was motivated by data availability and the relevant literature on high research performance. Because of the panel nature of the study, a logistic regression model with fixed effects was used. Three models were run for the three classes of top performers (the top 10 %, top 5 %, and top 1 %; Table 8). To check the assumption of a lack of multicollinearity in the vector of independent variables, the inverse matrix correlation method was used. The analyses were conducted for each period and each class of top performers (see more details in Electronic Supplementary Material).

### *4.3. Regression results*

In the regression models, we analyze which predictors change the probability of entering the classes of top performers (Table 8). The model for the top 10 % identifies several important predictors that increase the odds of success: gender, academic age, and various collaboration measures. Men have significantly higher chances (39.2 % on average; Exp(B) = 1.392) compared with women, with a narrow confidence interval (1.354–1.432) emphasizing the clear effect of gender on high publishing productivity. Each additional year of academic age increases the chance of success by 7.2 % on average (Exp(B) = 1.072), with a narrow confidence interval (1.071–1.073), underscoring the stable effect of academic age. Average team size, international collaboration rate, and general collaboration rate also positively impact the likelihood of being a top performer, with respective Exp(B) values of 1.009, 1.010, and 1.005. Each additional percentage point of the general collaboration rate (in the 0–100 range) increases the probability of success by an average of 0.5 %. The international collaboration rate matters even more, and each additional percentage point increases the chance of success by 1 % on average, all other things being equal. A one-unit (one coauthor) increase in the average team size increases the chance of success by 0.9 %.

**Table 8**
Logistic regression with fixed effects estimates (fixed effects: periods and disciplines), odds ratio estimates of membership in the class of the top 10 %, 5 %, and 1 % (in terms of productivity).

| Top performers type | variable | Exp(B) | LB | UB | Pr(> \|z\|) |
|---|---|---|---|---|---|
| The top 10 % | Academic age in period | 1.072 | 1.071 | 1.073 | <0.001 |
| | Average team size in period | 1.009 | 1.007 | 1.011 | <0.001 |
| | International collaboration rate in period | 1.010 | 1.009 | 1.010 | <0.001 |
| | General collaboration rate in period | 1.005 | 1.004 | 1.005 | <0.001 |
| | Gender: Male | 1.392 | 1.354 | 1.432 | <0.001 |
| | Institutional research intensity: Rest | 0.953 | 0.925 | 0.981 | 0.001 |
| The top 5 % | Academic age in period | 1.074 | 1.072 | 1.075 | <0.001 |
| | Average team size in period | 1.010 | 1.007 | 1.013 | <0.001 |
| | International collaboration rate in period | 1.011 | 1.010 | 1.011 | <0.001 |
| | General collaboration rate in period | 1.006 | 1.005 | 1.007 | <0.001 |
| | Gender: Male | 1.597 | 1.536 | 1.661 | <0.001 |
| | Institutional research intensity: Rest | 0.970 | 0.932 | 1.010 | 0.141 |
| The top 1 % | Academic age in period | 1.081 | 1.078 | 1.084 | <0.001 |
| | Average team size in period | 1.010 | 1.005 | 1.015 | <0.001 |
| | International collaboration rate in period | 1.013 | 1.012 | 1.014 | <0.001 |
| | General collaboration rate in period | 1.009 | 1.006 | 1.012 | <0.001 |
| | Gender: Male | 2.252 | 2.055 | 2.469 | <0.001 |
| | Institutional research intensity: Rest | 0.905 | 0.832 | 0.984 | 0.020 |

For the top 5 %, the effects of these predictors are even more pronounced. Being a male (Exp(B) = 1.597) and academic age (Exp(B) = 1.074) continue to play significant roles, and international collaboration rate (Exp(B) = 1.011) and general collaboration rate remain influential. The average team size has a slightly higher effect. For the top 1 %, the predictors' effects reach their peaks. Gender (Exp(B) = 2.252) and academic age (Exp(B) = 1.081) show the strongest influence, with international collaboration rate and general collaboration rate continuing to have an impact. Average team size remains a relevant factor.

Interestingly, not working in research-intensive institutions shows an only modest negative association for two classes of top performers (Exp(B) = 0.953 for the top 10 % and Exp(B) = 0.905 for the top 1 %; for some reason, it is not statistically significant for the top 5 %). This may suggest that, against expectations, institutional factors are much less critical for high productivity compared with individual characteristics. Being outside of the group of the 10 most research-intensive institutions decreases the chance of success only by about 5 % for the top 10 % and by about 10 % for the top 1 %.

The strength of influence of all significant parameters increases as we move up the top performance scale, highlighting the increasing impact of gender and academic age on higher productivity levels. Most notably, gender increases its impact as we move up the top performance scale. Being male as a predictor of high publishing productivity increases its influence significantly: Exp(B) increases from 1.392 for the top 10 % to 2.252 for the top 1 %. The more selective the class of top productivity is, the higher the chances for men (and the lower the chances for women) to enter it, all other things being equal. The precision of these estimates decreases for more selective classes of top performers, which is reflected in wider confidence intervals, yet all predictors remain statistically significant.

We have analyzed the top 10 % productivity class, and the values of the fixed effects for each period steadily decrease over time (Supplementary Table 5), which implies that, in successive periods, it becomes increasingly difficult to belong to this class. There is increasing competition in academia because of an overall increase in the number of publications. The highest values in disciplines such as ENG, MED, and PHARM imply that there is lower competition to belong to the top 10 % (compared with the average for the model for all disciplines). The lowest values observed for CHEM, EARTH, MATH, and PHYS indicate high competition in these disciplines.

For all top performer classes, the values of fixed effects decrease as the classes of top performers narrow from the top 10 % to the top 1 %, meaning that it is relatively more difficult to get into the narrower classes of top-productivity scientists (Supplementary Table 5). At the same time, for each class of top performers, the values of fixed effects decrease with each successive period, meaning that the competition is increasing over time.

In each successive period, entering each class of top performers is more difficult, with the most difficult class to enter being the top 1 % in the 2016–2021 period. The downward trend in fixed effects over successive periods indicates increasing competition in academia and increasing publishing productivity requirements to be among the most productive scientists.

In terms of disciplines, the direction of the influence of fixed effects is relatively constant for each top performer class. In some disciplines (e.g., ENG, MED, and PHARM), the positive values of fixed effects for all classes of top performers are maintained. In these disciplines, it is relatively easier to become a member of the top performer class (compared with the average for the models for all disciplines). In contrast, in other disciplines (e.g., BIO and MATH), negative values of fixed effects are consistently maintained, meaning that, in these disciplines, it is relatively more difficult to become a top performer. The values of fixed effects decrease with the narrowing of the classes of top performers for CHEM, EARTH, MATH, and PHYS (so that more selective classes of top performers are associated with increasing relative difficulty of entering them).

In practical terms, some disciplines are more competitive and require much more scholarly effort compared with the average effort in these disciplines (as measured in terms of publications) to be in the top-productivity classes; other disciplines are less competitive, requiring less effort to achieve the same high productivity status.

## 5. Discussion and conclusions

We were interested whether the role of top performers in the Polish science system is stable or increasing over time. We were wondering whether there would be increasing concentration of research activities (leading to increasing concentration of published research output) by top performers. Our results show that, in statistical terms, the role of top performers as knowledge producers has been surprisingly stable over the studied three decades, The role of top performers emerges from our research as fundamental today as it was 10, 20, and 30 years ago.

The instability of the national context for doing academic science over the past three decades comes in sharp contrast to the stability of productivity patterns we have found in the very same, fundamentally volatile period for society and its universities. The past three decades in Poland have been a period of intensive social and economic transformations that have been accompanied by several waves of higher education, research funding, and research assessment reforms (Antonowicz et al., 2020; Donina et al., 2022). In this volatile environment, for 30 years, the contribution of top performers to the national research output has been perfectly stable.

In the current research, the contribution of top performers to the total national research output in Poland was examined from temporal (1992–2021) and disciplinary (15 STEMM disciplines) perspectives. Three classes of top performers (the upper 1 %, 5 %, and 10 % of scientists in terms of publishing productivity) were examined separately within the disciplines and five 6-year periods. We examined a large population of all internationally visible—through Scopus-indexed publications—Polish scientists (between $N$ = 23,166 scientists in the first period and $N$ = 93,092 scientists in the last period) and a large sample of top performers (between $N$ = 2400 in the first period and $N$ = 9337 in the last period for the top 10 %).

Our methodological approach had four major attributes: It was longitudinal, relative, classificatory, and journal prestige-normalized. The longitudinal nature of our study allowed us to track scientists over time for the entire study period or as long as they were publishing. The study's relative nature allowed us to rank scientists by productivity for each period. Its classificatory nature allowed us to use a binary distinction between the three classes of top performers and the rest of the scientists. Finally, its prestige-normalized nature allowed us to use a citation-based vertically stratified structure of the global journal system (provided by Scopus) in productivity computations.

Our results show that the contribution of top performers to the total national research output is surprisingly stable over time: Although the social and economic world experienced powerful transformations and academic science was probably undergoing the biggest shifts in its recent history, Polish top performers from 1992 to 2021 played a structurally similar, fundamental role. The top 1 %, on average, account for 10 % of the national research output (which can be termed the 1/10 rule), and the top 10 % account for almost 50 % of the national research output (which can be termed the 10/50 rule), with significant disciplinary variations.

We could have expected the scale of their potential contributions based on previous cross-sectional, small-scale, and survey-based research (Kwiek, 2016; Kwiek, 2018a), but a comprehensive longitudinal approach covering three decades was not possible without access to the complete Scopus raw dataset.

Our research is not cross-sectional, with results only being valid for a single point in time (e.g., for two to four years), much like in traditional academic profession surveys. With panel data, long-term processes can be examined. The proportions of contributions made by top performers hardly change over time, even if shorter time periods are used; the processes of concentration of research output in the hands of the most productive scientists are not observed.

A possible explanation is that the distribution of highly productive scientists among subsequent cohorts of Polish scientists is relatively stable over time, and this may be related to cumulative advantage processes in an underfunded science system as well as to the distribution of individual motivation, high-quality doctoral training, innate ability, and possibly luck (Allison & Stewart, 1974; Stephan, 2012). The population of Polish scientists may be structurally divided into its small and much more productive (motivated, well-trained, able and lucky) segments and the rest. The proportions of output produced by scientists from subsequent percentiles of productivity seem relatively stable over time: Both the upper 1 % and upper 10 % have their stable shares.

Top performers classes emerge in the current research as heterogeneous classes in terms of gender and age (or publishing experience) groups. Women in the top 10 % are responsible for about 35–37 % of all publications produced by women. In addition, the concentration of output is higher for men in the top 10 % who are responsible for more than 50 % of all publications produced by men.

The concentration of output increases with age groups so that midcareer and late-career top performers—scientists with at least 20 years of publishing experience—from the top 10 % class are responsible for more than 60 % of all publications in this age group.

There are significant disciplinary variations, with the highest concentration being found in PHARM and the lowest in MATHS. A generally lower concentration for MATHS (in the range of 35 % to 40 %) may indicate that mathematicians are less vertically stratified in terms of publishing productivity: They contribute more equally to the total output in mathematics compared with other disciplines.

To quantify the overrepresentation of men in the top performers classes from cross-disciplinary and temporal perspectives, we constructed a special index (the RPI for men). Men are overrepresented in all classes of top performers, and their overrepresentation increases when moving up the high-performance scale. In terms of academic age groups, the RPI for men generally decreases with academic age: The youngest men are overrepresented the most, and the oldest men are overrepresented the least, indicating generational changes in the Polish science sector. In other words, across all periods and for the three classes of top performers, women are underrepresented the least among the oldest academic age groups and underrepresented the most among the youngest academic age groups.

Our econometric models lead to several interesting conclusions (logit generalized linear models with fixed effects, fixed effects being periods, and academic disciplines). First, the model for the top 10 % identifies three important predictors that increase the chances of membership in the top 10 %: gender, academic age, and publishing patterns. Men, on average, have significantly higher chances of success (39.2 %); each additional year of publishing increases the chances by 7.2 %; and an increase in the international collaboration rate and in general collaboration rate in a given period by one unit increases the chances by 1 % and 0.5 %, respectively. Thus, the chances of success are, on average, substantially higher for men with long publishing histories and involved in international research collaboration.

Second, for two (the top 10 % and the top 1 %) classes of top performers, working outside of the group of 10 research-intensive institutions plays a modest and negative but statistically significant role. The role of this institutional characteristic compared with the role of individual characteristics in the model is relatively weak. Although this finding is somehow counterintuitive: Our expectations were that top performers should be found predominantly in top, research-intensive institutions, so this may be caused by the binary classification of institutional research intensity in which a very small and productive PAS sector (about 4.8 % of scientists in 2023) was clustered together with the rest of institutions.

Third, a fixed effects analysis indicates that, over time, it is increasingly difficult to enter the top 10 %, which testifies to the increasing competition in academia. It is relatively more difficult to enter more selective top performers classes, with the most difficult class to enter being the top 1 % in the latest period of 2016–2021. However, the odds ratio estimates vary significantly between disciplines. Some disciplines emerge as more competitive than others, requiring more scholarly effort (measured in terms of publications) to enter top-productivity classes compared with the average effort in these disciplines.

Finally, in the current research, there are several limitations related to the data and methodology. Our sample includes all Polish scientists in the STEMM disciplines whose research articles (and papers in conference proceedings) have been indexed in the Scopus dataset from 1992 to 2021. As a result, scientists without such defined outputs are not included in the sample. However, it is extremely rare for scientists working in STEMM disciplines in Poland not to publish in Scopus-indexed journals.

A set of limitations is related to several proxies used, including the classification of disciplines, academic age, journal percentile ranks in Scopus, and institutional research intensity. We have used a commercial classification of disciplines (Scopus ASJC classification) rather than the various national classifications applied in Polish national research assessment exercises over the past 30 years. The ASJC classification offers a fine-grained level of 333 unique fields of research; however, we used 27 subject-level fields (the 2-digit rather than the 4-digit Scopus classification codes) because the number of observations per period by gender for unique fields of research would make our analyses meaningless. Hence, a more fine-grained variability within our disciplines cannot be properly taken into account. Our Laboratory of Polish Science includes the year of birth (coming from the national registry of Polish scientists) for a large number of scientists in our sample but not for all of them. Therefore, we have decided to use academic age (the time elapsed since the first publication until 2021) as a proxy of biological age (for correlations about the two age types, see Kwiek & Roszka, 2022). In all our productivity computations, we have used the most recent Scopus CiteScore percentile ranks (2023) because the percentile ranks for the whole period of study are not publicly available.

Finally, from several approaches to select a class of Polish research-intensive institutions, our choice was to use the most recent ministerial classification (research-intensive institutions vs. the rest), with the current vertical stratification of institutions—as in the case of the current vertical stratification of Scopus journals above—applied retrospectively for the whole study period. With no standardized reliable data for the past, in both cases, we assumed a higher stability of journal prestige and of research intensity of institutions over time than may have been the case in 1992–2021. These limitations reflect the trade-offs that had to be made to meaningfully study Polish top performers from longitudinal and cross-disciplinary perspectives.

As part of our current research agenda, we are also examining the role of the various classes of top performers globally and across 38 OECD countries; however, this research is in its early stages of data extraction and preprocessing.

*This paper is accompanied by* Electronic Supplementary Material *available online.*

## Funding information

We gratefully acknowledge the support provided by the MEiN NDS grant no NdS/529032/2021/2021. We also acknowledge the assistance of Kristy James and Alick Bird of the International Center for the Study of Research (ICSR). We want to thank Dr. Lukasz Szymula from the CPPS Poznan Team for his support with Scopus data acquisition and analyses.

## Data availability

We used data from Scopus, a proprietary scientometric database. For legal reasons, data from Scopus received through collaboration with the ICSR Lab cannot be made openly available.

## CRediT authorship contribution statement

**Marek Kwiek:** Writing – review & editing, Writing – original draft, Validation, Supervision, Resources, Project administration, Methodology, Investigation, Funding acquisition, Formal analysis, Conceptualization. **Wojciech Roszka:** Writing – review & editing, Writing – original draft, Visualization, Validation, Software, Methodology, Investigation, Formal analysis, Data curation, Conceptualization.

## Declaration of competing interest

None.

## Supplementary materials

Supplementary material associated with this article can be found, in the online version, at doi:10.1016/j.joi.2024.101595.